# Design evolution for the Wide-field Spectroscopic Telescope

Will Saunders[a*], Corentin Cudennec[b], Roland Bacon[b], Philippe Dierickx[b], Roelof de Jong[c], David Lee[d], Gaston Gausachs[e]

[a]Astralis, Australian Astronomical Optics, Macquarie University (Australia). [b]Université Claude Bernard Lyon 1, CNRS, CRAL (France). [c]Leibniz-Institut für Astrophysik Potsdam (Germany). [e]UK Astronomy Technology Centre (United Kingdom).[g]Advanced Instrumentation Technology Centre, ANU, Canberra (Australia).*will.saunders@mq.edu.au

**ABSTRACT**

WST is proposed as the next large ESO project to follow ELT, combining Multi-Object Spectroscopy and Integral Field Spectroscopy. Each mode offers order-of-magnitude gains over current systems, and each also presents unprecedented design challenges, both separately and in combination.

For large MOS systems, both science performance and spectrograph costs vary steeply with the delivered image quality, so exceptional delivered image quality is paramount. But the 12m aperture and 2 degree field give WST an etendue larger than Rubin, larger that LAMOST, and larger than all other existing MOS telescopes combined; while the IFS, segmented primary and windy site all add additional constraints. Hence finding designs with good image quality is challenging. Eventually, 3-lens Forward Cassegrain designs with loss-less ADC were developed in variants mostly differentiated by M2 diameter. The lowest technical risk design, with the smallest M2, was selected as the baseline, with wind-shake control a primary driver. However, other designs have better as-designed image quality, and their perceived risks may diminish as the system design and underlying technologies mature.

For IFS mode, delivered image quality is just as crucial, but this is achieved through additional optics and NGS GLAO over the 3'x3' field. The challenges come from (a) transferring the F/3.4 Forward Cassegrain focus to a fixed focus under the telescope; (b) a requirement that the 3'x3' field be selectable from a 13' diameter FoV without repointing the telescope; (c) including a suitably conjugated mirror for GLAO correction; (d) doing all this with minimised vignetting and surface count. Various designs were explored; the baseline design has a field-selecting pick-off at telescope focus, combined with large reimaging optics at Nasmyth, giving an F/28.5 fixed IFS focus.

## [1] INTRODUCTION

The Wide-Field Spectroscopic Telescope is proposed as the next large ESO project to follow ELT. It combines Multi-Object Spectroscopy with Integral Field Spectroscopy, with each mode intended to give an order-of-magnitude gain over current systems. This already tightly constrains the overall telescope design: the 12m aperture is driven by the need to be more sensitive than Subaru+PFS and VLT+MUSE, while the 2° FoV is driven by the need to survey the whole accessible sky in 5 years. The resulting etendue is larger than the Rubin telescope, larger than LAMOST, and larger than all other existing MOS telescopes, combined.

The Delivered Image Quality (*DIQ*) is paramount in both modes (Section 2). The *DIQ* is given by the seeing distribution, as modified by the Facility Image Quality (*FIQ*, i.e. the telescope and dome contributions). The *FIQ* has many components, mostly but not always harmful - ground-layer turbulence from the building; dome and mirror seeing; outer scale effects; wind-shake; optical aberrations, both as-designed and as-built; distortion changes due to differential refraction; guiding errors and fast-guiding benefit; and flexure and vibration of both telescope and instrument during the exposure. This paper is mostly concerned with the MOS-mode 'Zemax Image Quality' (*ZIQ*), that is the as-designed telescope aberrations. It becomes increasingly challenging to achieve good Z*IQ* with increasing etendue: a larger FoV means more resolution elements across the focal surface, while a larger aperture becomes increasingly constrained by maximum available lens and mirror sizes, which are in turn limited by existing technologies, and also weight, stiffness and stability. *ZIQ* would normally be expected to be the dominant contribution to the *FIQ* in a wide-field telescope, and to do justice to the seeing. All MOS telescopes with fibers also need to be 'effectively pupil-centric', that is that the Numerical Aperture (NA) of the telescope on-axis must also suffice at other field angles, or nearly so. This effect becomes increasingly important as etendue increases, since spectrograph design difficulty and cost become paramount.

Simultaneous fiber-fed MOS and IFS use is also unprecedented. The scientific advantages are obvious, but so are the technical obstacles: the telescope focal position and speed for wide-field fiber-fed MOS are not at all what one would choose for IFS, and relaying the IFS beam inevitably constrains the MOS design. The Forward Cassegrain design is preferred on MOS Z*IQ* grounds (Section 4), and at least this puts the focus close to the elevation axis, very helpful for transferring the IFS beam to an elevation-independent position at Nasmyth. But doing this unfavorably constrains both the focal height above M1, and the back-focus below the Wide-Field Corrector (WFC).

There are additional constraints:
- the site will be windy (not yet fixed, but all options are in northern Chile, close to Paranal, and sharing its windiness), greatly constraining M2 size to allow fast-steering wind-shake compensation.
- the segmented primary, with the need to view all of M1 for phasing, even those parts never used for collecting useful sky light; this constrains the use of M2 to balance vignetting and pupil-centricity (Section 5), and also complicates the baffling.
- a decision was taken in 2024 to adopt a conservative telescope design, despite the huge leap in etendue. In particular, a constraint was imposed that M1 should if at all possible be a pure conic.

## [2] *DIQ*, SURVEY SPEED, AND SPECTROGRAPH COSTS

### Spectrograph Etendue

The 'spectral etendue' per MOS fiber (fiber solid angle x telescope area x spectral resolution elements) is twice as large as existing systems (PFS/4MOST/DESI), rendering the MOS spectrograph design very challenging. But also, there are 30,000 fibers, an order-of-magnitude increase on existing systems, giving a total spectral etendue ~30 x PFS/4MOST/DESI. So there is a tension in the spectrograph design: it is intrinsically difficult, but also needs to be simple, compact, affordable, and readily manufactured, installed and maintained. Both the costs and risks are large for the spectrographs, and hence anything that eases the spectrograph design is very precious.

### Lloyd's law[1]

For a telescope of fixed size and for fixed spectral resolution requirements, the *DIQ* impacts the spectrograph design via the fiber size and hence the required camera size and/or speed, and hence the spectrograph difficulty, throughput, volume and weight, cost and risk. In practice, camera size is strongly constrained by lens and detector availability, while speed is constrained by optical considerations. In the WST regime, both dioptric and catadioptric solutions are possible. But the former are at their limits in terms of size and speed; while the latter are much larger, more complex and more expensive. For now, we assume the classic scaling law for optics (cost $\propto$ diameter$^{2.5}$), to get

$$\textbf{MOS spectrograph costs} \propto \boldsymbol{DIQ}^{\,2.5} \qquad (1)$$

But *DIQ* also has a dramatic effect on faint object survey speed, via sky noise, and camera efficiency. For unresolved faint sources, the noise is proportional to *DIQ*, while the spectrograph throughput declines with camera speed, especially if we have to pass from dioptric to catadioptric solutions. As a crude first scaling, we assume

$$\textbf{Survey Speed} \propto \boldsymbol{DIQ}^{\,-2.5} \qquad (2)$$

This means that for faint objects,

$$\textbf{Spectrograph cost per unit survey speed} \propto \boldsymbol{DIQ}^{\,5} \qquad (3)$$

The index can be debated, but is clearly very large, and certainly above 4. For bright sources, the scaling is not much less steep - the S/N scales less steeply with *DIQ,* but the spectrograph issues are similar or even worse.

So, although *DIQ* requirements for spectroscopy are often assumed to be lower than for imaging, this is not true for a massive cost-limited MOS survey such as WST. Excellent *DIQ* is likely to be the only way to render the survey affordable.

1 Argument originally given by Jamie Lloyd

**The Schlegel bonus**[2]

Happily, there is another factor that aids in the endeavor of an affordable massive MOS system. That is, the simple observation that because the survey speed is faster in better seeing, **most data has sub-median seeing**. And if the facility is designed to take advantage of good seeing, then this effect is accentuated, leading to a further improvement in the median data *DIQ* and the survey speed, and lower spectrograph costs as well. Figure 1 quantifies this argument, showing that improving the *FIQ* from say 0.5" to 0.2" improves the survey speed by 33%, while reducing fiber size by 28%, increasing the speed per unit spectrograph cost by a factor ~2.5, the same as the change in *FIQ,* so *FIQ* is comparably as important as throughput. Of course, a better facility costs more money, and only part of the *FIQ* is amenable to improvement via a better telescope, and gains in *ZIQ* that come at the cost of e.g. increased wind-shake may be counterproductive. But the potential gains and losses are very large, and merit careful consideration.

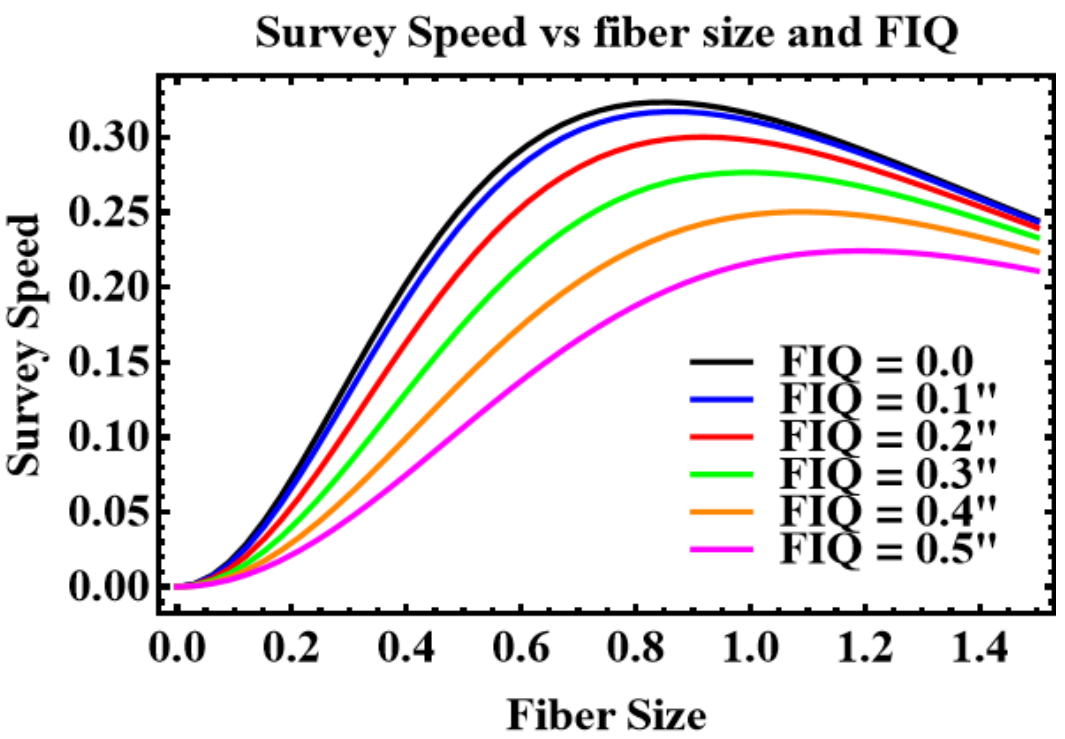

Figure 1: Overall survey speed versus fiber size and *FIQ*, for face-on $R_e$=0.1" galaxies, averaged over the Paranal seeing distribution, and ignoring overheads. Improving the *FIQ* by a factor 2.5, from 0.5" to 0.2", reduces the optimal fiber size by 28%, while improving the survey speed by 33%. For the scalings in Equations (1), and (2), this translates to a Speed/ Spectrograph Cost improvement by the same factor of 2.5.

Note that a *FIQ* ~ 0.32" would give an optimal fiber size of 1" (the baseline assumed value). This then represents the nominal *FIQ* 'budget' before the fibers become undersized.

The survey speed is normalised so that the total galaxy flux is 1 unit, and the sky noise 1 unit per arcsec$^2$.

**The *WIQ* and its pitfalls**

The nominal WST image quality requirement is defined as 'WST *IQ*' (*WIQ*) > 90%. This means a <10% loss of point source injection efficiency into a 1" fiber in 0.5" seeing at Zenith at 500nm, just due to *ZIQ*. At first sight this appears to be a demanding specification. But, the aperture value is large compared with the seeing value (the optimum Petrosian diameter is a factor just 1.5 times the seeing for a Moffat $\beta$ = 2.5 profile). And counter-intuitively, the better the nominal seeing value used in the *WIQ*, the *less* stringent the requirement becomes.

Figure 2 shows the the cumulative encircled energy as a function of radius, with and without the *ZIQ*. It also shows the Petrosian diameters, and indicates that a 90% *WIQ* implies a ~15% increase in optimal fiber diameter. But a 15% increase in fiber diameter imposes an increase in spectrograph costs of 40%, and a similar penalty in overall survey speed, and hence a 2-fold increase in the spectrograph cost per unit science, as compared with a perfect telescope. And this is before accounting for all the other *FIQ* contributions. Hence, a 90% *WIQ* is much less good than it sounds.

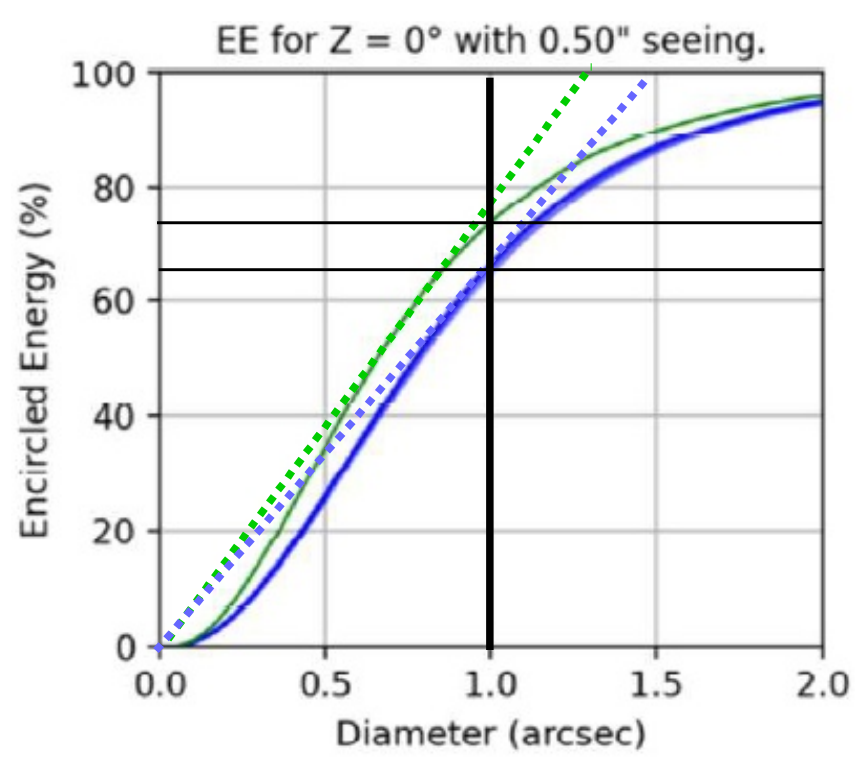

Figure 2: Cumulative encircled energy vs fiber size for WST at Zenith. Green is 0.5" seeing, blue is when convolved with the telescope *ZIQ*. The *WIQ* is the ratio of these values at 1".

It happens that the Petrosian diameter (which maximises the S/N for faint point sources) is given by the tangent point from the origin to these curves (dashed lines). This optimum diameter increases due to *ZIQ*, by much more than the loss of *WIQ* - a 90% *WIQ* leads to a ~15% increase in the diameter. The increased fiber size decreases the survey speed by ~40%, and increases the spectrograph costs by ~40%, with a combined effect of doubling the spectrograph cost per unit survey speed.

2 Argument originally given by Dave Schlegel

**IFS Image Quality**

For the IFS, image quality is at least as important as for MOS, defining the sensitivity and limiting depth of the instrument. There is a strong desire to have an image quality at least as good as VLT+MUSE, which benefits from GLAO via an adaptive M2 [1]. A benefit of the optical train necessary to transfer the IFS from Cassegrain focus to a fully fixed position is that it is easy enough to arrange for a deformable mirror of reasonable size and AOI, and conjugate to the desired height of ~100-200m above M1. A 6' FoV provides adequate Natural Guide Stars over ~90% of the sky. This means that the *ZIQ* of the MOS telescope is less relevant to the IFS, since monochromatic aberrations can be corrected in the IFS optics train, and chromatic aberrations by powering the windows that are needed in any case. The IFS image quality is considered in detail in Dierickx et al (2026) [2, D26].

**Image Quality metrics and scalings**

*DIQ*, *FIQ* and *ZIQ* are always quoted as FWHM in this paper. Where other *IQ* metrics are quoted (rms radius, or $d_{80}$), they can be scaled as follows: for a Gaussian distribution (a fair approximation for aberrations), the FWHM is 1.664 x the rms radius, and $d_{80}$. is 1.52 x the FWHM; while for a Moffat $\beta = 2.5$ distribution (as assumed for Paranal), the FWHM is 1.251 x the rms radius, and $d_{80}$. is 2.454 x the FWHM.

## [3] WST HISTORY

WST started life as Bernard Delabre's 12m SpecTel design with 2.5° FoV [4,5] (Figure 3). It includes a coude focus for IFS use (but not simultaneously with MOS). The image quality is poor, unsurprising given the enormous FoV and classic (rather than Forward) Cassegrain design. The design includes the newly developed 'Loss-less ADC' [6,7,8], whereby displacement of a WFC lens and also M2 allows an ADC action with almost no degradation of the monochromatic image quality. This minimises the lens count and also allows an all fused silica WFC, and hence larger lens sizes and excellent UV/NIR performance.

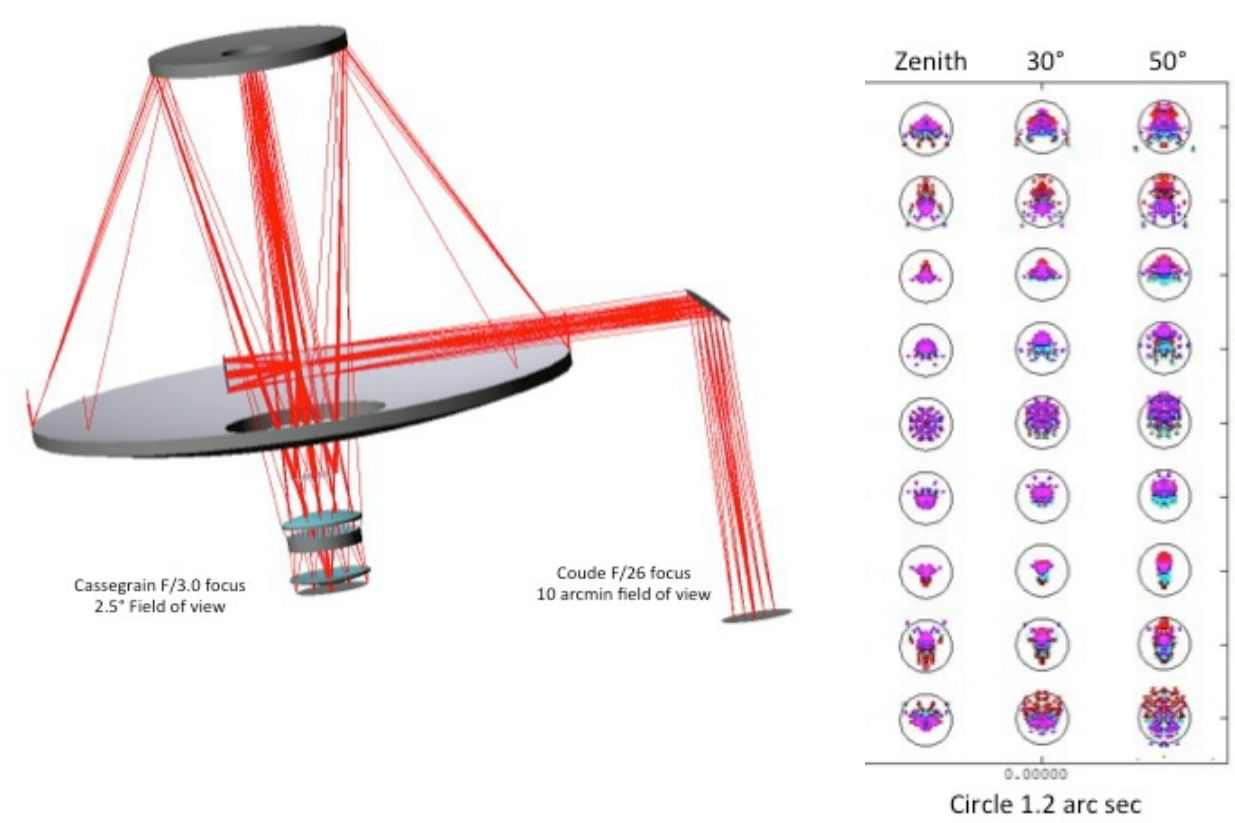

Figure 3:First SpecTel design. Note large M2, classic rather than Forward Cassgrain, poor MOS image quality (circles are 1.2"), dual (but not simultaneous) foci, and off-axis IFS focus.

The design was subsequently modified to a Forward Cassegrain design with FoV restricted to 2° (Figure 4) [9]. The design allows simultaneous MOS + IFS use, with the IFS focus on-axis, and with the 3'x3' IFS science field selectable from the IFS 13' FoV without telescope repointing. However, this comes at the cost of 20% IFS vignetting from the M3/M4 pick-off system, and a convoluted IFS optics train. The M2 size was 3.1m, but with the express desire to reduce this to 2.5m, to allow wind-shake correction at 3-5Hz. The image quality remained poor ($d_{80}$ >~ 0.7-0.8"). At this stage, the MOS design bore a striking resemblance to the Forward Cassegrain design proposed for MSE [10]. The MSE design had a somewhat smaller (1.52°) FoV, but with image quality 2-3 times better, suggesting that the WST *IQ* could be significantly improved.

In 2024 a three-mirror anastigmat 'Quad' design was briefly considered, with 1.8° FoV. The design was not adopted, due to issues of cost, complexity, FoV, throughput and pupil-centricity. Cassegrain designs with both 3 and 4 lens WFC were also considered in 2024, but the improvement from an additional lens was only marginal. At the start of the EU-Horizon study in early 2025, a design with good MOS image quality (i.e. much better than the seeing) had not yet been developed.

Figure 4. WST design presented at SPIE 2024 [4]. A Forward Cassegrain design had been adopted, which aids extraction of the IFS beam, and also optimises MOS image quality. But the MOS image quality is quite poor ($d_{80}$ >~ 0.7-0.8"), there is 20% IFS vignetting (due to M4) and the IFS train is convoluted.

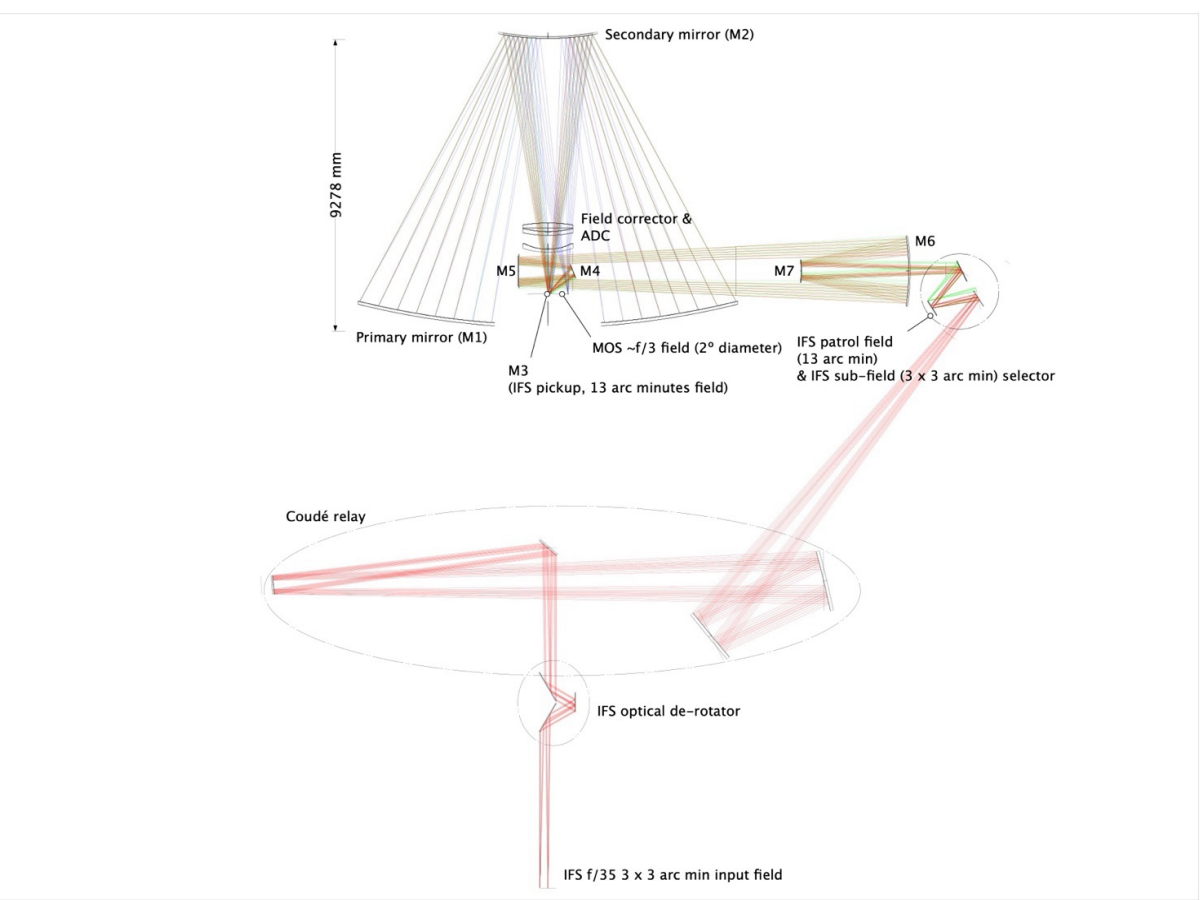

## [4] IN SEARCH OF BETTER IMAGE QUALITY

There are multiple ways to improve image quality, offering larger and smaller gains.

**Bigger M2:** 2.5m is very small for a wide-field 12m telescope - much smaller than Rubin (3.1m), and barely larger than MUST (2.4m, with barely half the M1 size). But a larger M2 has reduced stiffness, and so is harder to use to control wind-shake. This is discussed in more detail in D26 [2]. A 2.7m M2 seems plausible (i.e. has an eigenfrequency > 50Hz, 10 x the desired correction frequency), either in SiC or light-weighted Zerodur, and allows a design with better *IQ*, as discussed below. Larger than that would likely require adopting a different way of controlling M2, eg the pneumatic supports of VISTA, or the hybrid actuators of TNO (with potential stroke ~100um) [11]. The latter might also allow some modest GLAO correction of ground-layer, dome and mirror seeing.

In principal, it might be possible to correct for wind-shake correction elsewhere. A solution with wind-shake corrected by movement of L3 was investigated, the optics worked well enough, but the prospect of displacing 400kg of glass by up to 1.5mm and at up to 5Hz (accelerations 0.17*g* or 700N of force), without disturbing the rest of the telescope seemed an unacceptable risk.

Note that the increase in diameter required for improved *ZIQ* is quite modest - a diameter of 3.0m seems sufficient, and even 2.7m allows a good solution (Section 4).

Figure 5 shows the distribution of wind speed and direction for excellent and poor seeing conditions. Strikingly, there is very rarely excellent seeing at Paranal when it is windy. The degree to which the telescope design should be optimised for windy conditions then needs very careful consideration, accounting for both the observational requirements and the expected wind-speed vs seeing distribution.

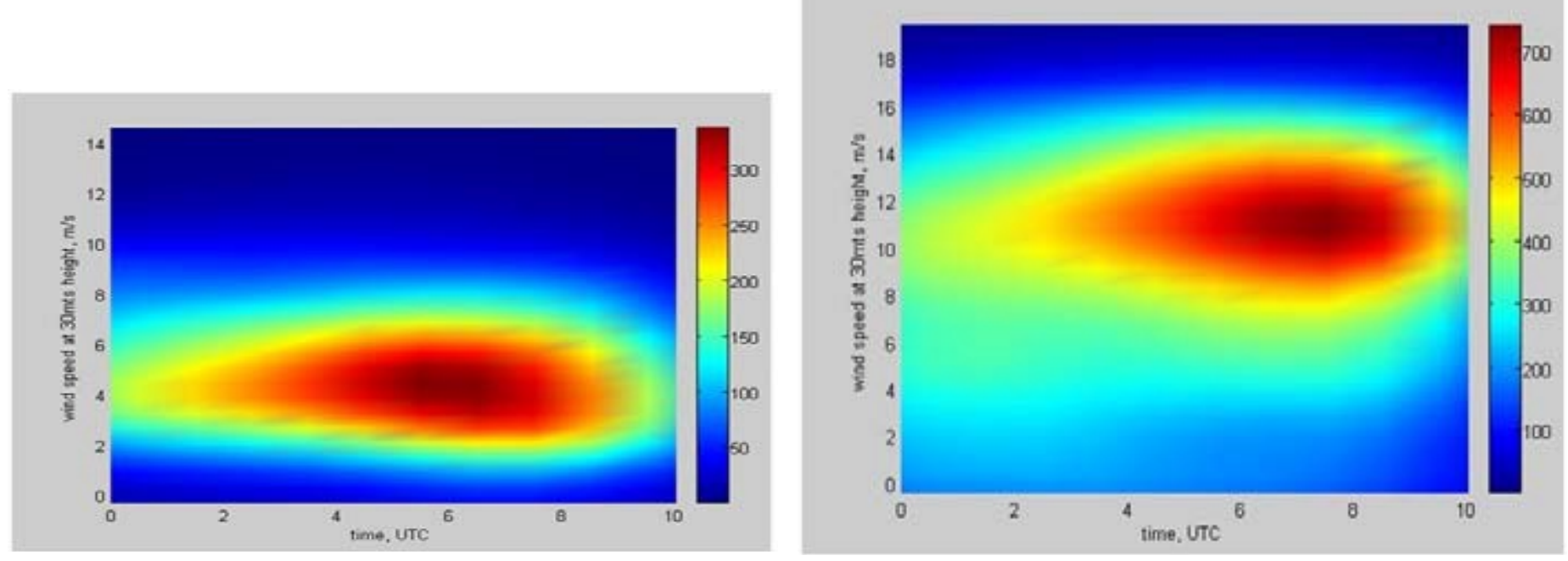

Figure 5. Paranal distribution of (a) superb seeing (FWHM<0.5") and (b) poor seeing (FWHM>1.5") vs wind speed, showing that the best seeing rarely occurs above wind speeds 8m/s, and almost never above 10m/s. Note that the Y-axes have been rescaled to be the same for both plots. Adapted from Navarrete [12].

**Aconic M1:** An aconic M1 (like Ruben) improves *IQ*, but only in combination with a larger M2, for reasons not yet understood. Only a very small aconic term (amounting to ~10µm asphere at the edge of the M1) is required. This option is discussed in detail below. M2 already has an aconic departure, of similar size, in all the designs.

**More WFC lenses:** Designs with 4 and even more lenses were investigated. However, the gain is only incremental, and must be balanced against lost throughput and increased cost and weight. In any case, the large back-focus required to extract the IFS beam heavily constrains this idea. Eventually, adequate 3-lens designs were found and adopted.

**Larger WFC lenses:** Larger WFC lenses always help the *IQ*, because they can be further from focus. L1 is already undersized at 1.6m diameter, versus 1.7m actually required to avoid vignetting. If larger lens sizes become available, image quality (and/or vignetting) would be improved.

**Greater WFC lens asphericity**: L2 is the strongest asphere, with Zemax demanding ~60mrad aspheric slope, vs a reasonable limit of 25-30mrad, as used for e.g. the 4MOST and DESI corrector lenses. However, since L2 is biconcave, both surfaces can be aspherised. There still needs to be a transmission test for homogeneity, but homogeneity is not normally a major concern for fused silica lenses. The two surfaces also need to be adequately coaxial, but the tolerance on this seems acceptable (~100um). This offered the largest *IQ* improvement of all the options listed here.

**Telescope speed F/3.1-F/3.2:** The optimal speed for best *IQ* (in arcsec) appears to always be in this range for a 12m Cassegrain design with 2° FoV, at least one with three lenses limited at 1.7m and no WFC vignetting. This speed is also close to optimal for minimising Focal Ratio Degradation while avoiding cladding losses in normal (NA=0.22) fibers. However, the effect is modest, and the multiple other constraints normally override it.

**Allow IQ degradation at large field angles**. The baseline MOS positioner layout is hexagonal, which allows efficient tiling on the sky and also gives sufficient room for wavefront sensing and aquisition/guiding cameras. This means that the fully filled area of the focal surface extends only to the inscribed circle at 0.866° radius (Figure 6); demanding good image quality beyond this radius is counterproductive if IQ consequently declines at smaller radii. For a survey telescope such as WST, the integration times are likely set by the sensitivity at a rather low quantile of the positioners - eg 4MOST uses the 10% quantile. It happens (and seems not accidental, given that 4MOST also has a hexagonal positioner layout) that this is almost exactly the fraction of positioners beyond the inscribed circle. So assuming that WST follows the same logic, 0.866° is the relevant radius for determining the overall MOS survey speed.

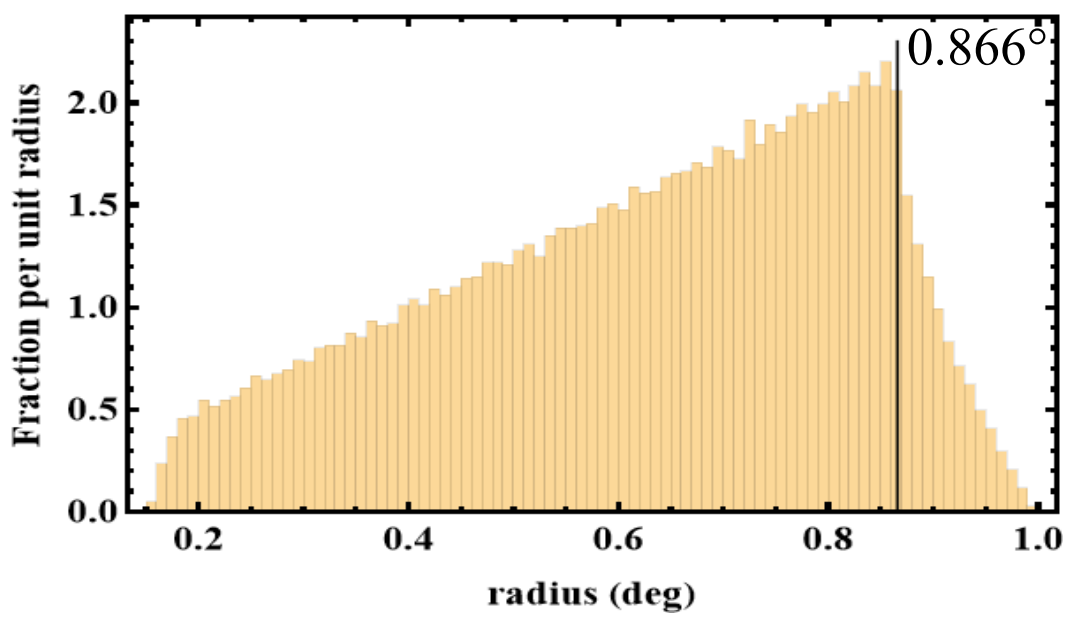

Figure 6: Radial distribution of field angle for the MOS positioners, assuming a hexagonal layout with curvilinear modules [13]. Beyond 0.866°, the positioner density drops markedly, and the total positioner fraction beyond 0.866° is just 9.5% of the total.

**Creative use of vignetting at M2 to create an 'effectively pupil-centric' focal surface[3]:** Light can only be lost once, so there is (almost) no additional loss to vignet light that would in any case be normally lost through collimator overfilling. So it makes sense to clip this light at M2. Modest vignetting allows increased chief ray misalignment for off-axis fields, while minimising the M2 diameter and improving IQ (since it is the fastest beams that are clipped). This is explained in more detail in Figure 7.

In any case, for an etendue-limited telescope such as WST, there is not much point in in trying to capture all the light. Figure 8(b,c) shows the clipped and unclipped pupil areas, showing that a 5% clipping of M1 diameter (and hence 10% loss of etendue) leads to just 5% light loss. However, the requirement that M2 be sized large enough to allow all of M1 to be visible for phasing - even parts not normally wanted for science - restricts this idea, and increases the diameter of M2, just when we are trying to minimise it. Even for the IFS, the pupil will be masked to circular, ~12m diameter, versus the actual mirror maximal diameter of 12.6m, forcing M2 to be 5% oversize with no benefit for IFS and not much for MOS.

---

3 The idea derives from the PFS wide-field corrector, brought to our attention by Paulo Spano

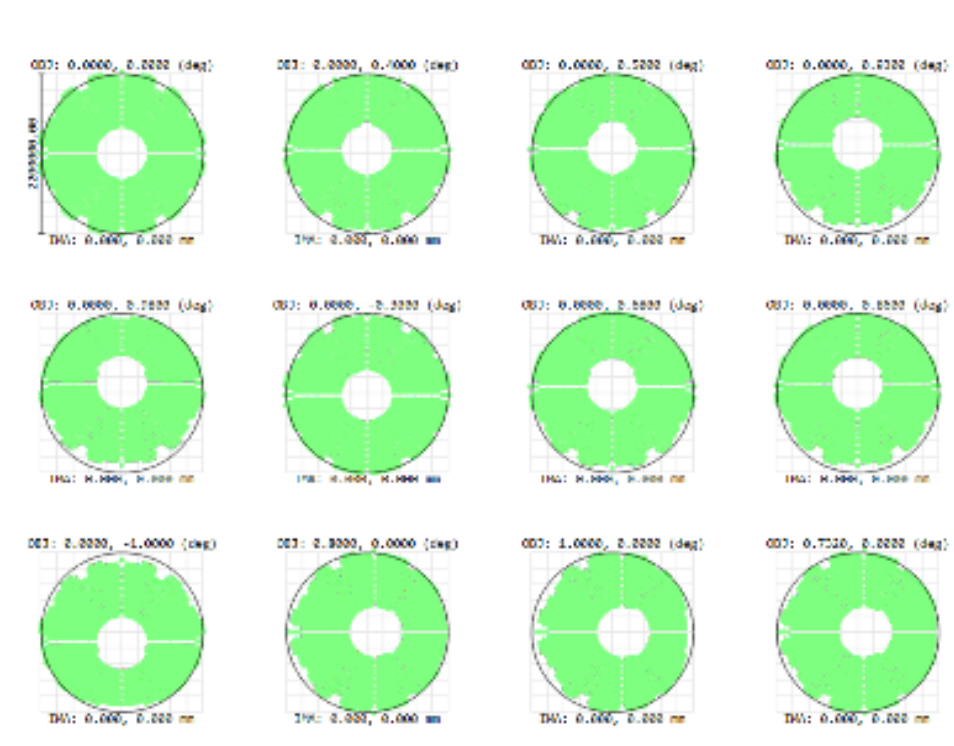

Figure 7: Far-field illumination of the fibers for the A27 design presented below. Fibers at various field positions are shown. The circle shows marginal rays at F/3.24, corresponding to the nominal telescope pupil of 12m on-axis. The rays clipped by M2 ensure that the Numerical Aperture of the light off-axis does not exceed the maximum on-axis value. This make the design 'effectively pupil-centric'. The idea apparently comes from PFS (though both the formal non-pupil-centricity and vignetting are much more extreme for PFS than for these designs). Note that there is also vignetting by L1; this is the flatter cut on the same side as the M2 vignetting at extreme field angles.

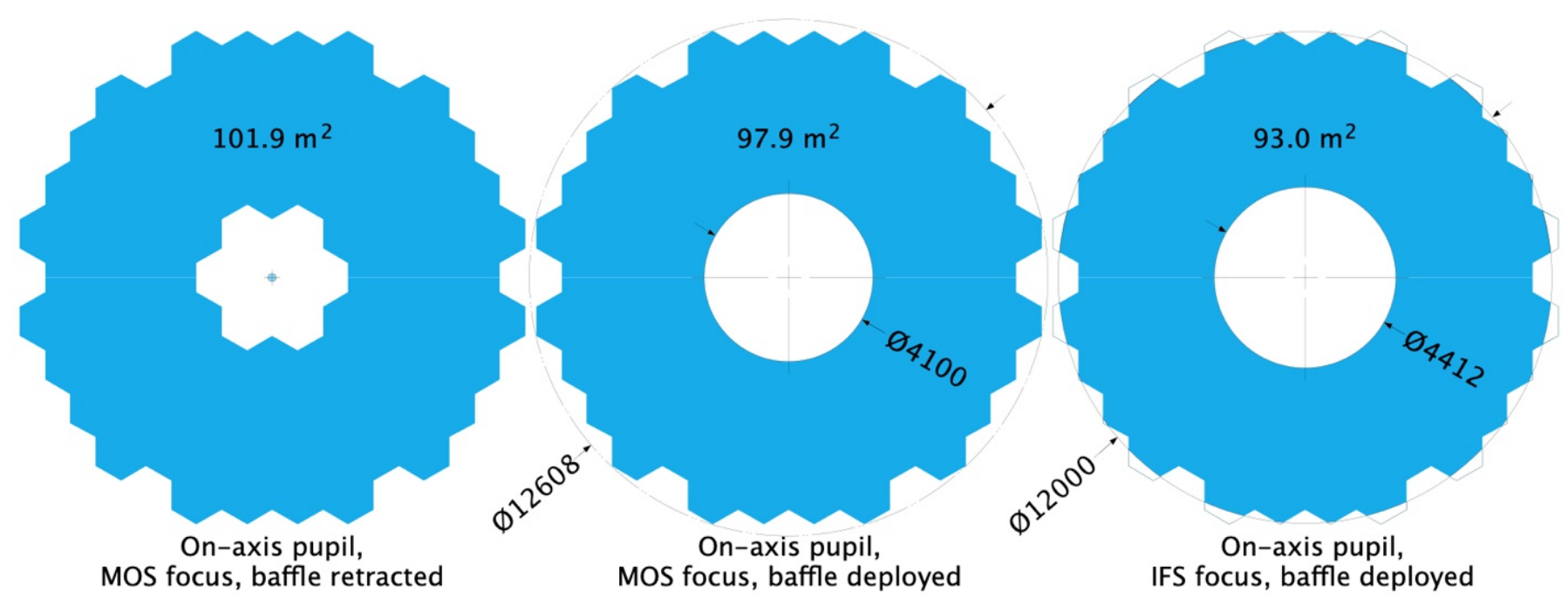

Figure 8. On-axis telescope pupil in different modes, showing the small loss of area (5%) for a larger reduction in etendue (10%) between a pupil of 12.6 and 12m.

## [5] A27 DESIGN

The A27 design is a forward Cassegrain design, with speed F/3.24 for a nominal 12.0m pupil. M1 is segmented, with 12.6m full diameter, speed F/0.975, $k$ = -1.090, with a 6th order polynomial term amounting to 10um at the outer edge of M1. M2 has diameter 2.7m, the smallest size that was found to not impose a significant *IQ* penalty (earlier versions had M2 of 3.0m and 2.9m, with slightly better *IQ*). It has $k$ = -5.16, with an additional 6th order polynomial term amounting to 6um at the edge. There is a 3-lens all-fused-silica corrector, with lens diameters 1.6m, 1.5m, 1.5m, with aspheres on L1/2, L2/1, L2/2 and L3/2, of 30, 31, 31 and 6mrad. The Focal Surface (FS) is spherical[4], 1340mm diameter for the 2° FoV, plate scale 185um/" on-axis, RoC 7000mm. The FS is 1.4m above the M1 vertex and the back-focal distance is 1.5m. There is a 4.1m M2 baffle and a Cassegrain baffle to prevent skylight from arriving directly at the focal surface. Optimisation was performed over 0.37-1.6um. The full layout is shown in Figure 9.

The ADC action is provided by moving L3 in an arc 'following the curve' by up to 45mm at ZD=65°, with 0.9° tilt and 4mm axial motion. The axial motion introduces a small plate-scale change, which allows the overall differential refraction caused by the changing ZD while tracking to be almost halved [14]. A simple method for moving L3 with the required single degree of freedom was presented by Peter Gillingham [11], but flexures are clearly viable also, and have been adopted for the baseline design [2]. A much smaller motion of M2 (4.5mm lateral motion, 0.2° tilt, 0.14mm axial motion, all within comfortable limits of the hexapod required in any case) almost entirely compensates for the monochromatic aberrations (astigmatism, coma, focus) introduced by the L3 motion. The polychromatic *IQ* is limited by secondary spectrum; fused silica is the obvious lens material in this size due to availability, homogeneity, and transparency, but it has rather a poor partial dispersion match to the atmosphere. L3 is sized for Zenith use, so the L3 motion causes a small amount of additional vignetting, just along one edge of the FoV, at large ZDs.

4 The preferred FLEX positioners [13], selected after this design was undertaken, allow an axial motion, so a small improvement in *IQ* can be gained by allowing a non-spherical focal surface.

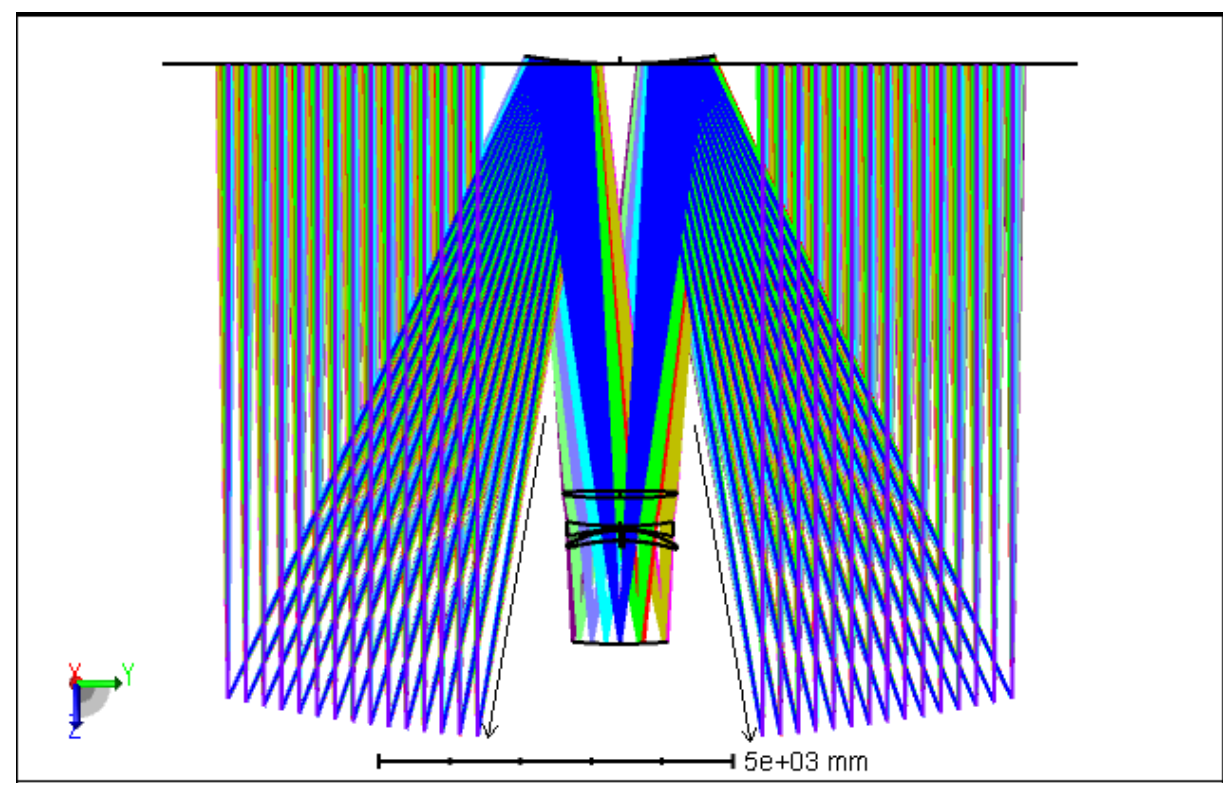

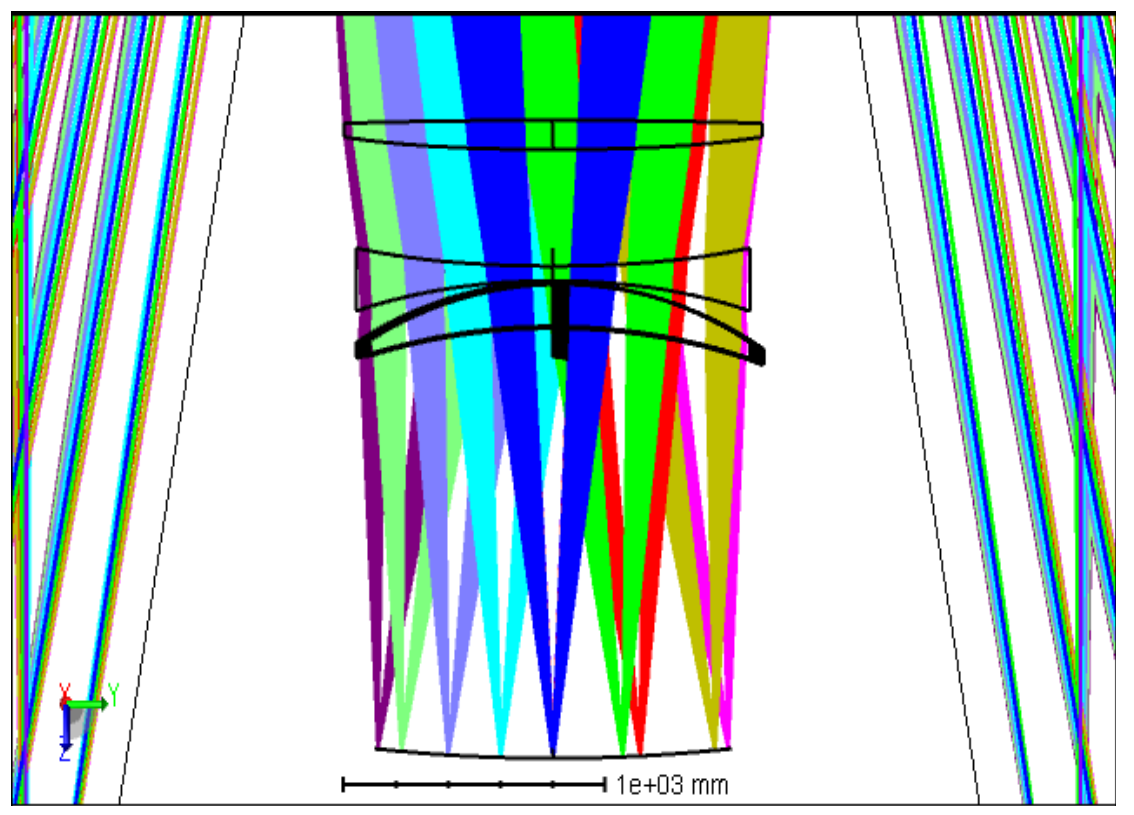

Figure 9: (a) overall layout and (b) close-up of the WFC, in both cases for the full range of ZD's 0-70°. The ADC action is provided by moving L3 as shown (essentially following the curve of the lens), with a maximum motion of 58mm.

**Image Quality**

The image quality is presented for ZD0 and ZD=45°. Figure 10 shows the spot diagrams, vs the expected fiber diameter of 1". Figure 11 shows the rms radius vs color and field position, showing monochromatic rms radii < 24μm (0.13") out to a field radius 0.866°. Figure 13 shows the polychromatic rms radius field map. Figure 14 shows the geometric enclosed polychromatic energy with radius, $d_{80}$<0.32" within 0.866°, degrading to 0.59" at 1°. In general, the *monochromatic* image quality degrades barely at all with zenith distance. This is what matters most for fiber spectroscopy, since virtually every spectrum has a critical wavelength where the required S/N is most challenging to achieve, and the fiber will normally be positioned for this wavelength.

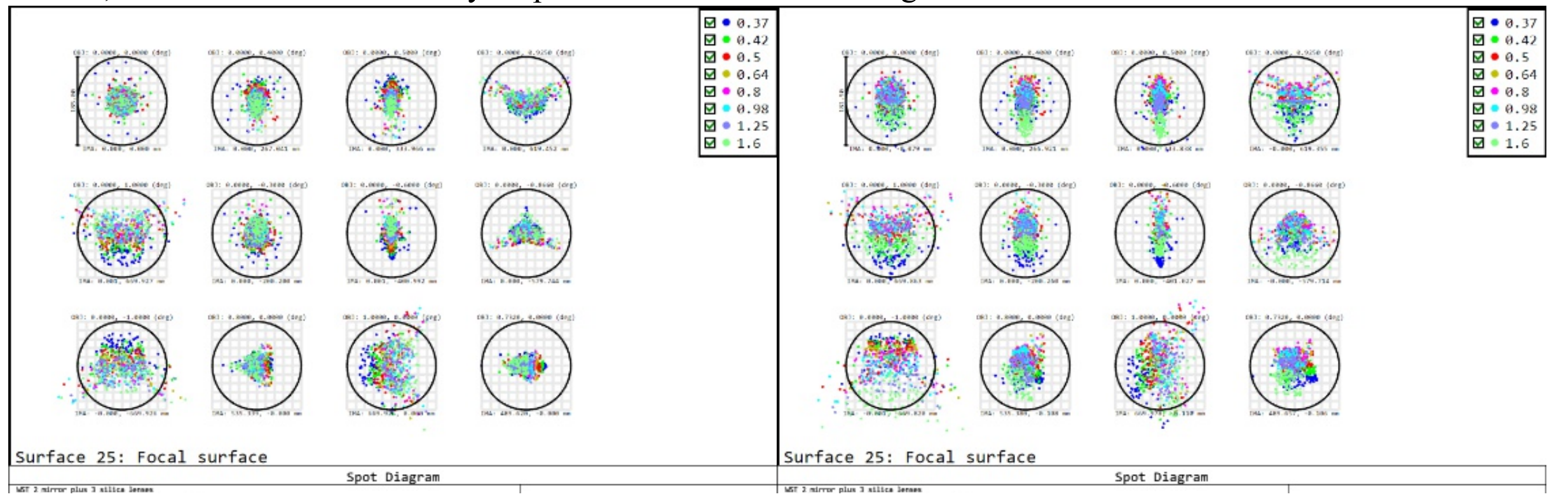

Figure 10. Spot diagrams for A27 at ZD0 and ZD=45°, for various field positions. Circle size is 1" (the expected fiber diameter). The image degradation at ZD=45° is almost entirely due to secondary spectrum.

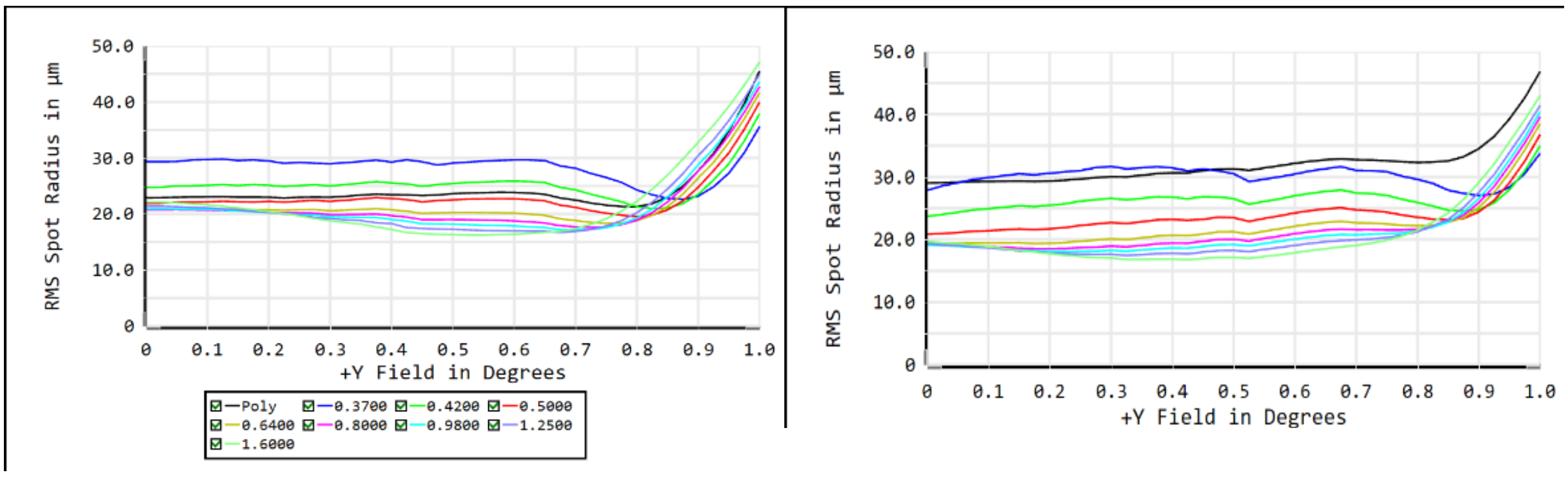

Figure 11. RMS radius vs color and radius for the A27 design at Zenith and at ZD=45°. At Zenith, the polychromatic rms radius is < 0.13" out to .866° and 0.24" at 1°. While the polychromatic image quality (black) degrades with ZD due to secondary spectrum, the monochromatic IQ degrades barely at all.

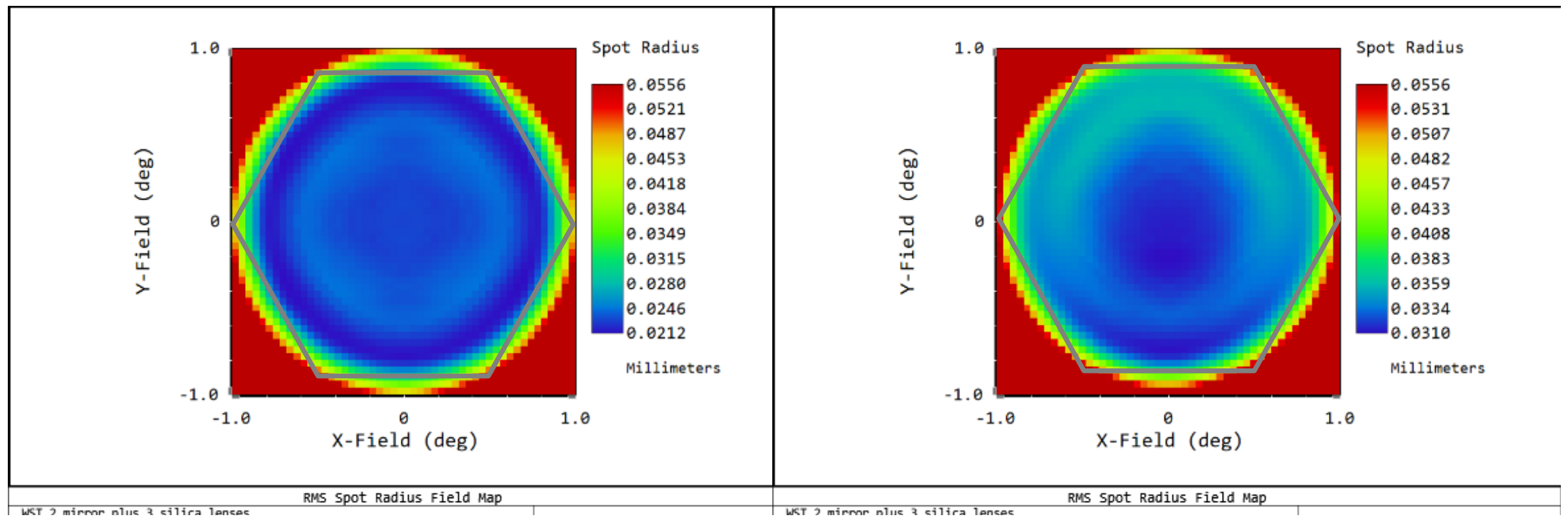

Figure 12. Rms radius maps at ZD0 and ZD=45°, showing the asymmetry in the *IQ* inherentt to the loss-less ADC design. The nominal positioner layout is also shown. Note that the intensity scale for the two plots is not identical.

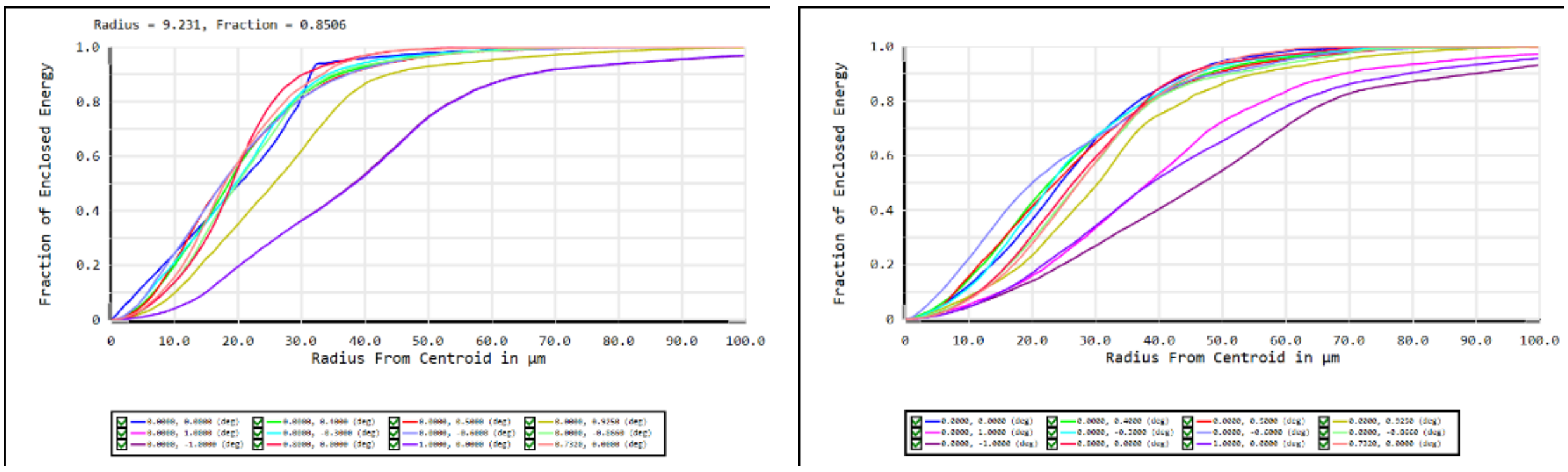

Figure 13. Geometric polychromatic encircled energy radius for various field positions, (left) for ZD0 and (right) at ZD=45°. At ZD0, the polychromatic $d_{80}$ is < 0.325" for all field radii < 0.866° degrading to 0.59" at 1°. The degration at ZD=45° is almost entriely caused by secondary spectrum (see Figure 11).

**Throughput, Vignetting, Pupil-Centricity**

The on-axis throughput losses are determined by the two mirrors, the 4.1m baffle, 6 air/glass surfaces, any underfilling of M1 beyond 11.1 m (see Figure 8). Off-axis, one must add vignetting at M2 and L1; at large ZD's, there is also vignetting along one edge of the field due to L3 displacement.

The vignetting caused by M2 and L1 are shown in Figure 14. The Y-axis is with respect to a 12.6m circular pupil.

Figure 14. Vignetting vs field radius at Zenith. The gentle decline is due to vignetting at M2; the sharper roll-off at large radii is due to L1 under-sizing. The normalisation is wrt a filled 12.6m circle; for a 12.0m pupil, the on-axis vignetting would be 0.86 (see Figure 8).

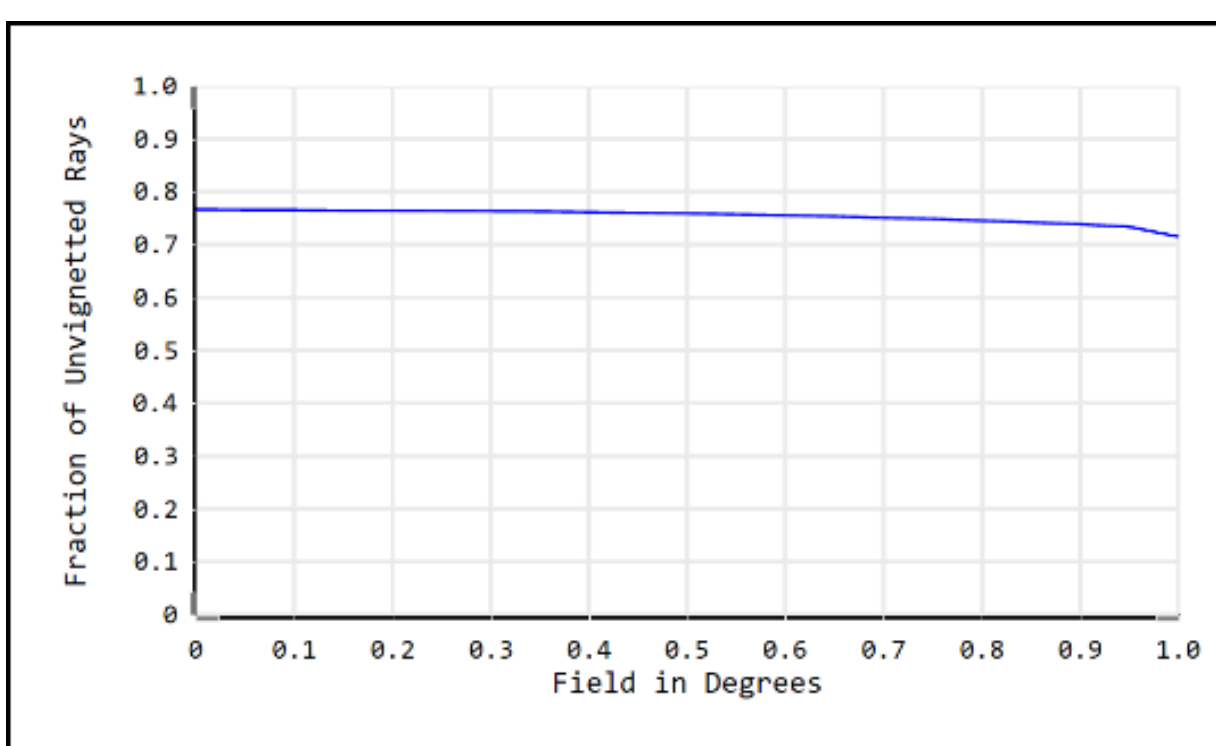

Pupil-centricity has already been discussed above, in Section 4 and Figure 7. The design is almost completely effectively pupil-centric, with < 1% of the light from a nominal 12m pupil at any field position and at any ZD falling outside the NA defined by the same pupil on-axis at Zenith.

**M1 aconicity**

The A27 design does not respect an imposed constraint that M1 should be a pure conic; it includes a 6th order polynomial term, amounting to 10.5µm at the outer edge. The Rubin primary includes a similar term, of similar size, so the benefit seems generic. If M1 is restricted to conic, then the *IQ* is degraded by ~15%.

**Distortion changes and differential refraction**

As the ZD changes, there is image motion due to the combined effects of differential refraction and ADC motion. The combined effect was minimised in the optimisation. The small axial motions allowed for M2 and L3 give a plate scale adjustment, which can reduce the quadrupolar distortion change caused by atmospheric refraction, at some cost in image quality. Figure 15 shows the distortion field changes between ZD0 and ZD=30°, amplified 1000-fold. The worst distortion changes are
55µm (0.30”) from ZD0 to ZD30°;
102µm (0.55”) from ZD30 to ZD5=0°;
139µm (0.75”) from ZD50 to ZD=60°
135µm (0.73”) from ZD60 to ZD=65°.

At large ZDs, these represent a significant compensation for differential diffraction (which is eg 1.26” from ZD=50° to ZD60°).

Figure 15: Combined effects of differential refraction and ADC action on the image positions betweem ZD0 and ZD=30°, magnified 1000-fold. The largest changes are 0.3". The ADC action allows a plate-scale change which partially compensates for the quadrupolar contraction caused by differential refraction.

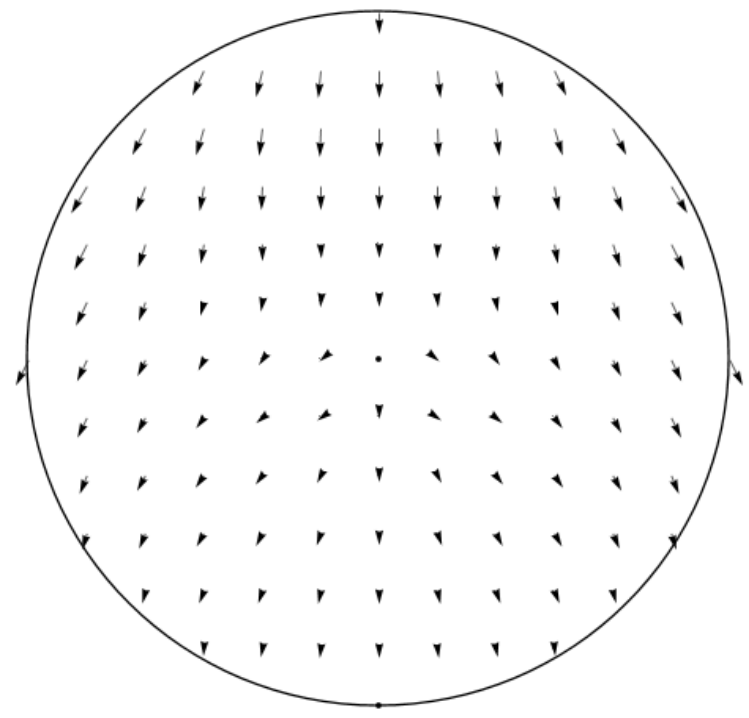

**Focal Surface speed, plate scale and curvature**

The speed of the telescope dictates the required collimator speeds, as modified by the Focal Ratio Degradation. For the latter, the fastest speed commensurate with negligible cladding losses is preferable, and the F/3.2 speed of the A27 design is close to optimal for normal (NA = 0.22 ± 0.02) broadband multimode optical fibers. However, the collimator design for the MOS-LR spetrographs is based on the slower (F/3.4) speed of the B25 design, and some readjustment of dichroic AOIs would likely to be necessary with a faster collimator.

The number of positioners is essentially limited by the cost of the required spectrographs, but for a fixed number of positioners, a larger plate scale is clearly always helpful. For the baseline positioner layout of 30240 actuators in a hexagonal layout [13], the required pitch for A27 would be 6.9mm, comfortably larger than the nominal minimum assumed pitch of 6.2mm. So the faster speed presents no show-stopping problems for the positioner layout.

Focal surface curvature leads to a subtle loss of effective sky coverage, for a hexagonally-packed positioner layout (like 4MOST, PFS, but not DESI). The number of fiber positioners and their combined patrol area is effectively determined by the area of the FS as seen in projection. But the actual FS area is larger, by a factor ~ $1 + \frac{1}{4} (r / R)^2$, where $r$ is the FS radius and $R$ its RoC. Depending on the patrol area, this effect may give actual coverage gaps, or may be hidden in the overlaps, but it cannot be avoided. In practice, the effect is very small - for A27, the loss amounts to just 0.2%.

## [6] B25 DESIGN

The B25 design adopts most of the features discussed in Sections 4 and 5, but has a pure conic M1, and M2 was limited to 2.50m clear aperture (but has subsequently increased to 2.56m to allow a view of all of M1 from a position on-axis, for phasing). The M1-FS separation is imposed as a constraint; the net effect is to reduce the telescope speed to F/3.395 for a nominal 12m pupil. The B25 design represents a remarkable achievement, in that the *IQ,* vignetting and pupil-centricity are all only modestly worse than the A27 design, despite the imposed constraints. It was selected as the baseline design in April 2025. The design has undergone modest evolution since that time, the numbers and plots shown below refer to the current version.

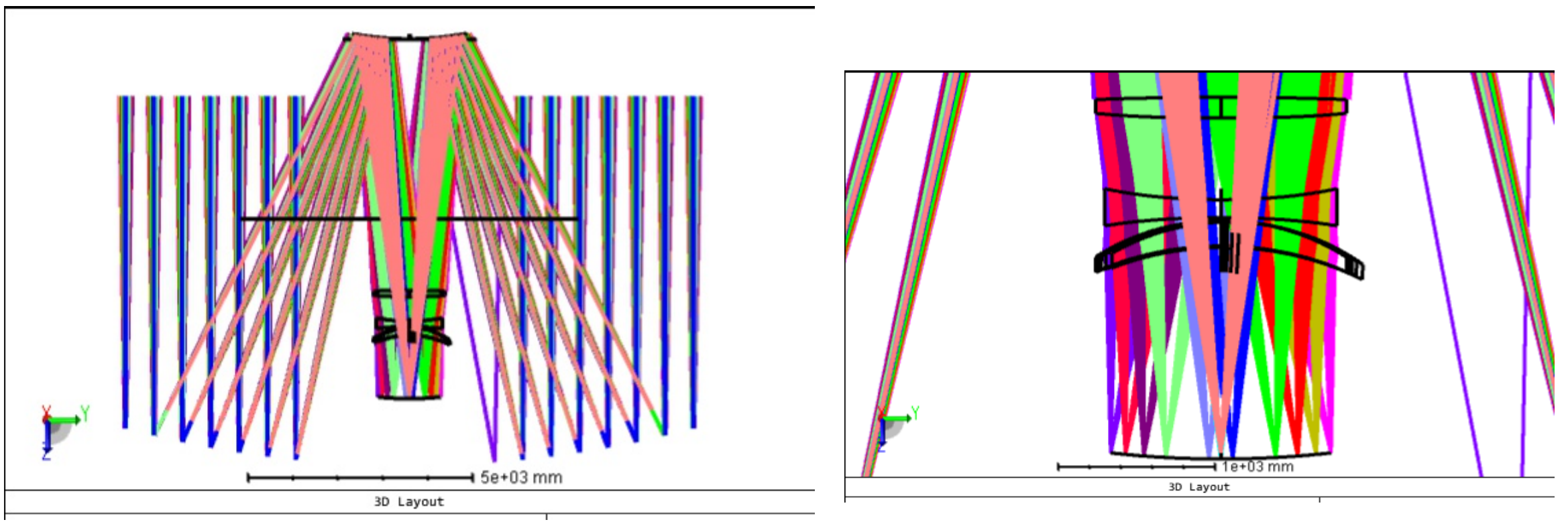

Figure 16. B25 design, showing ADC action via L3 and M2, in both cases for the full range of ZD's 0-70°. The maximum L3 motion is 118mm.

M1 is somewhat faster (F/0.95); The overall speed is F/3.395 for a 12m pupil; the plate scale is 195µm/", and the focal surface has diameter 1406mm with RoC 6021mm (so 20% more strongly curved than A27). The ADC action is again through a circular motion of L3, combined with a small motion of M2. The motions of L3 are twice as large as for A27 (e.g. 74mm at ZD=60°), meaning that there is increased vignetting on one side of the FS at large ZDs.

Spot diagrams are shown in Figure 17. The rms radius is shown in Figures 18 and 19. The polychromatic *IQ* is similar (and sometimes better) than A27 over about half the field area, but it degrades much faster at large field angles, being 1.5 x worse at and beyond 0.886° (0.19" vs 0.13"). The monochromatic *IQ* degrades little with ZD, but is rather variable with color, being poorer in the UV, surely due to the thicker corrector lenses. Geometric enclosed energy is shown in Figure 20.

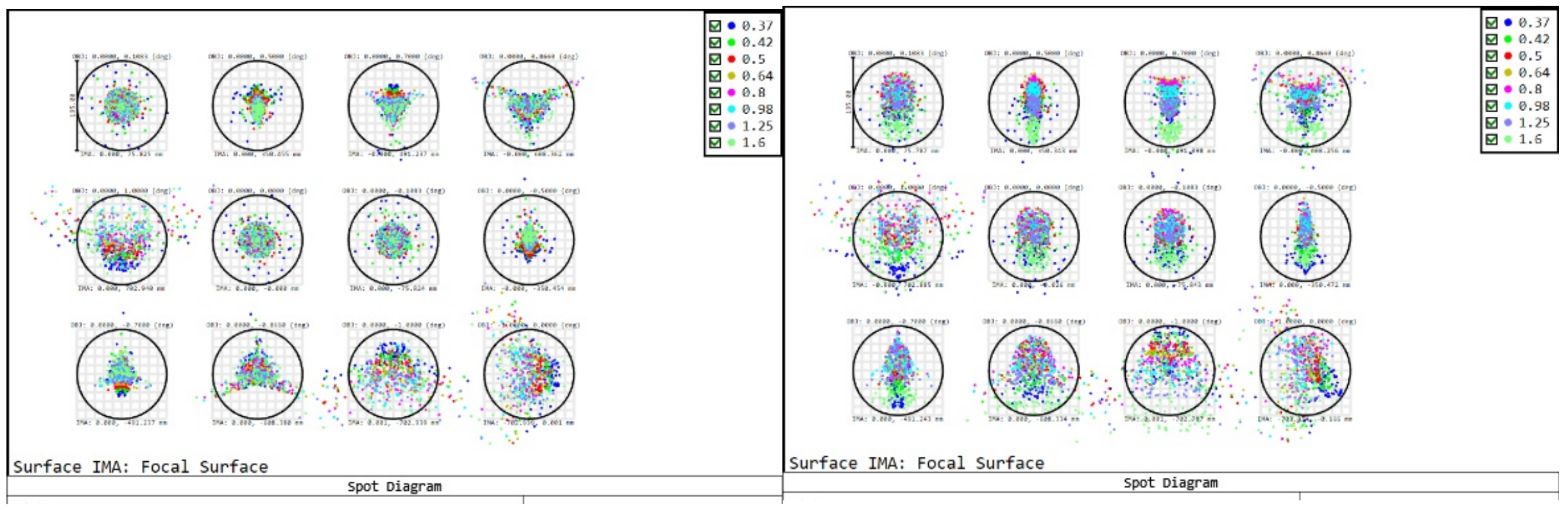

Figure 17. Spot diagrams for B25, at ZD0 and ZD=45°. *IQ* degradation is mostly due to secondary spectrum. Circle size is 1".

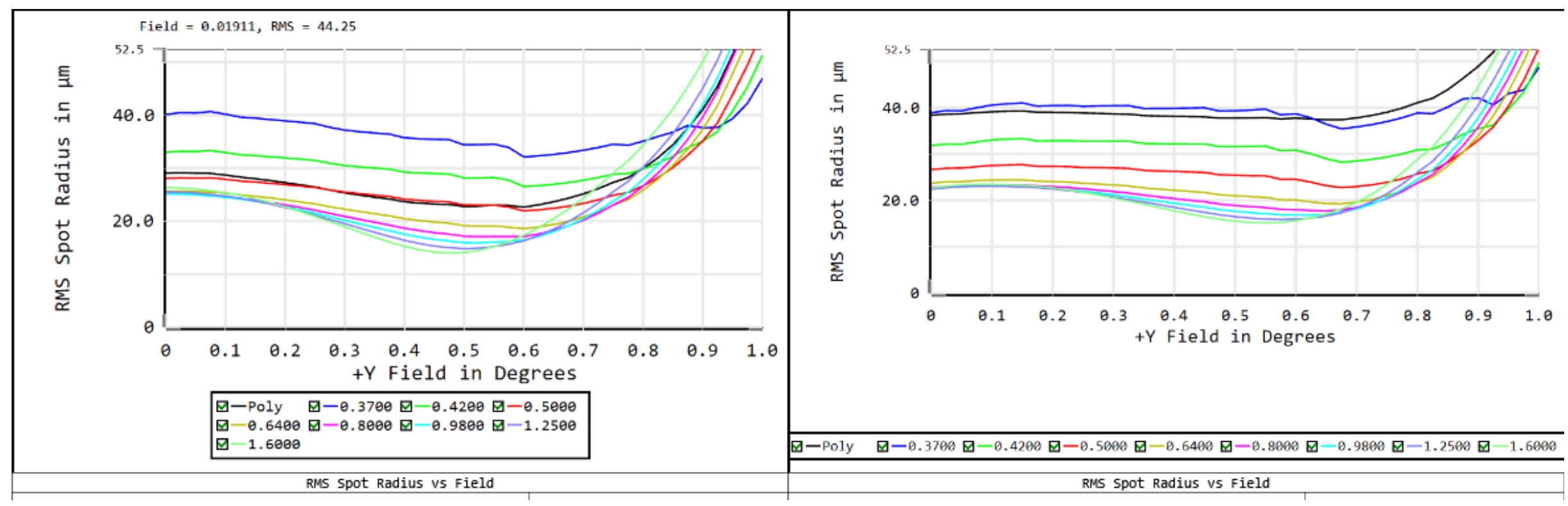

Figure 18. Rms radius vs wavelength and field radius for B25, for Zenith and ZD=45°. The Y-axis has been scaled to be the same in arcsec as for Figure 11. At ZD0, the polychromatic rms is < 0.15" within 0.79°, but increases rapidly thereafter, being 0.19" at 0.866° and 0.33" at 1°.

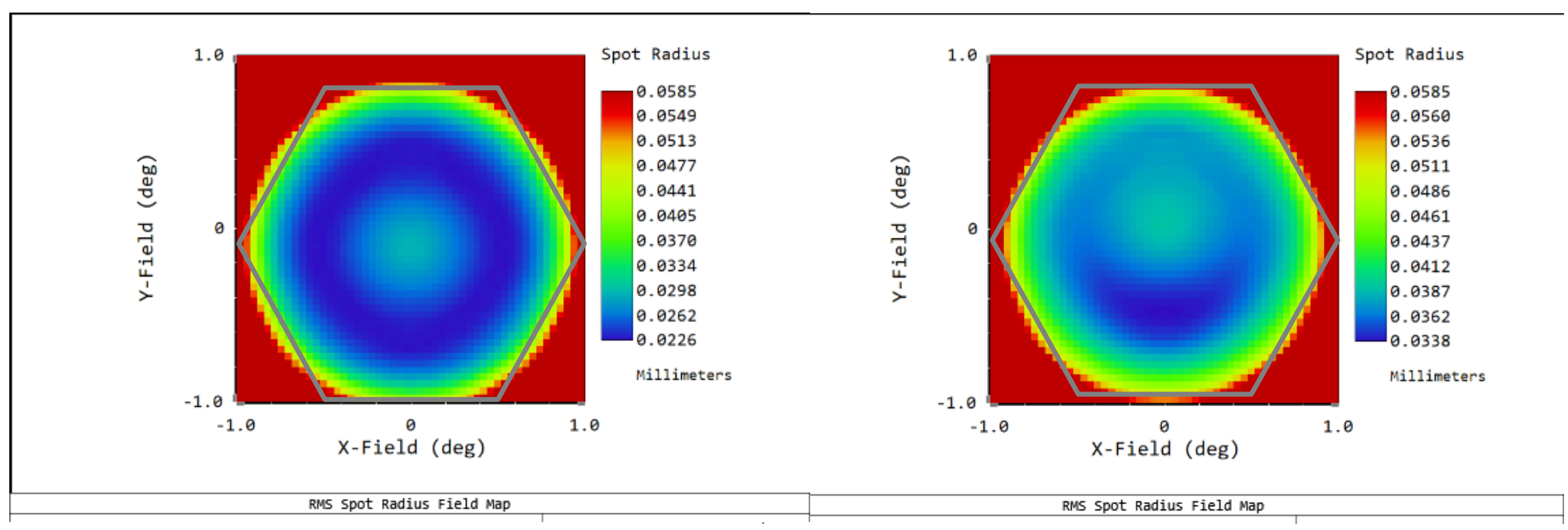

Figure 19. Rms radius field map for B25. The maximum intensity is 0.3", the same as in Figure 12. The nominal positioner layout is also shown. The minimum intensities are auto-scaled and are not identical.

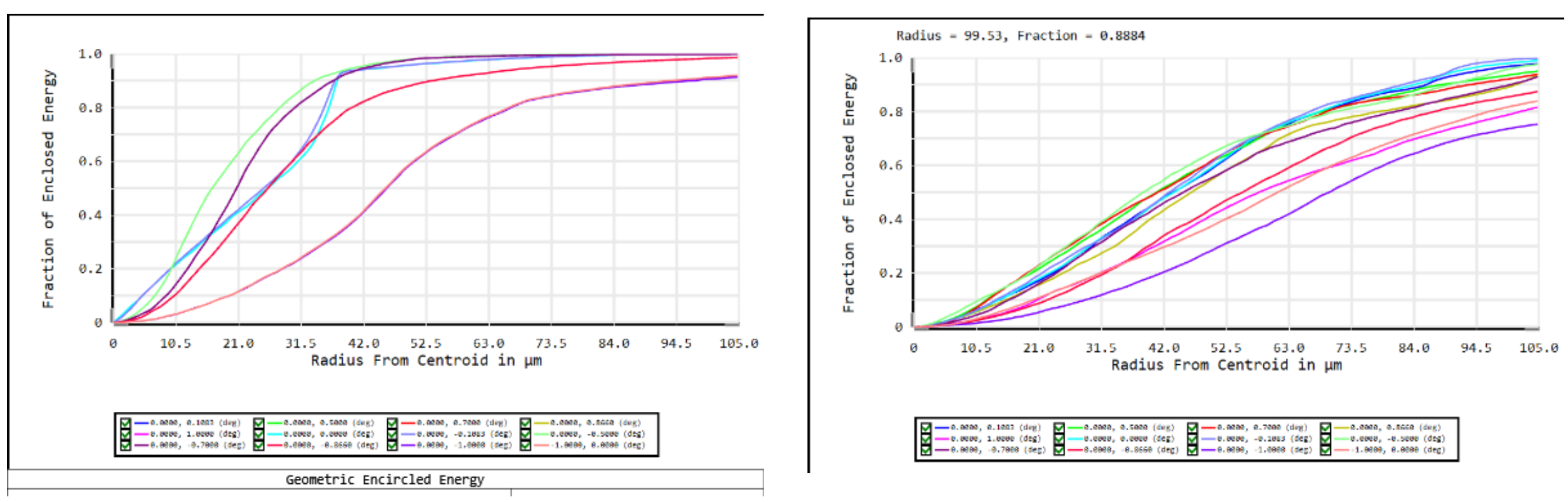

Figure 20. Geometric polychromatic encircled energy for B25, at ZD0 and ZD=60°. The X-scale is the same in (arcsec) as for Figure 13. Zenith $d_{80}$ is 0.41" within 0.866° and 0.68" at 1°.

### Throughput, vignetting and pupil-centricity

The throughput is essentially identical to A27. The vignetting is shown in Figure 21, it is also essentially identical, except at large radii (>0.9°), where the effect of L1 vignetting cuts in much faster. The pupil-centricity is shown in Figure 22 and is also essentially identical. The FRD will be somewhat worse due to the slower beam, leading to additional collimator overfilling losses crudely estimated at 1%.

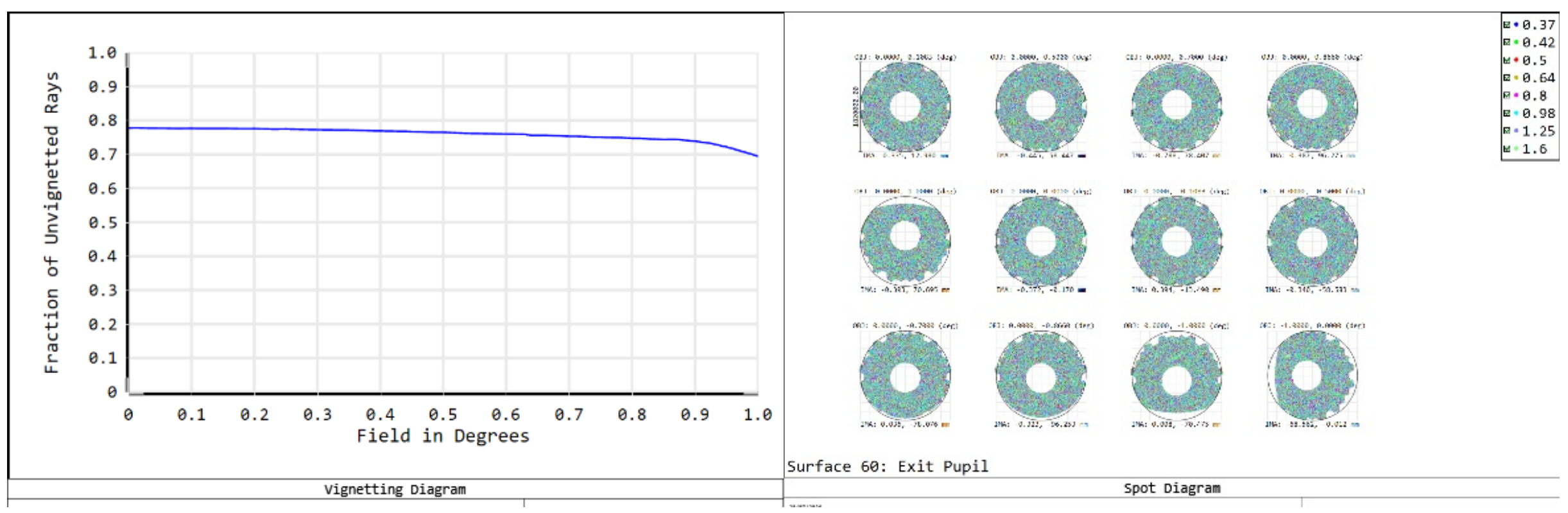

Figure 21 (left). Vignetting plot for B25. The throughput drops somewhat more strongly at large field radii than for A27, due mostly to vignetting at L1, as can also be seen comparing with Figures 7.

Figure 22. (right) B25 pupil-centricity, at Zenith. The circle shows the NA for a nominal 12.0m pupil at Zenith. Vignetting comes from M2 (the clipping of the beam arounf onr half of the beam) and L1 (the flatter cut on the same side).

### A27 vs B25 M1 shape

Interestingly, and for reasons not understood at this time, the B25 design benefits hardly at all from an aconic M1. This, combined with the smaller M2 and larger plate scale, was a significant factor in its adoption as the baseline design.

Clearly, the difficulty of manufacturing, testing, aligning and especially phasing segmented mirrors all increase with the curvature, the difference between the sagittal and tangential curvatures, and the rates of change of these quantities with radius. B25 is a pure conic, but has a smaller RoC. Figure 23 shows these quantities, for both A27 and B25, with the A27 curves plotted both with and without the 6th order aconic term (though the differences caused by this term are much too small to see, being always less than 0.04%). The differences between A27 and B25 are modest, but are always in the sense of a larger curvature, difference in curvatures, and derivitives of these quantities, for B25. Taking segmentation into account, a thorough review of the impact on fabrication meteology and in-situ M1 control would be necessary to understand and quantify the relative risks.

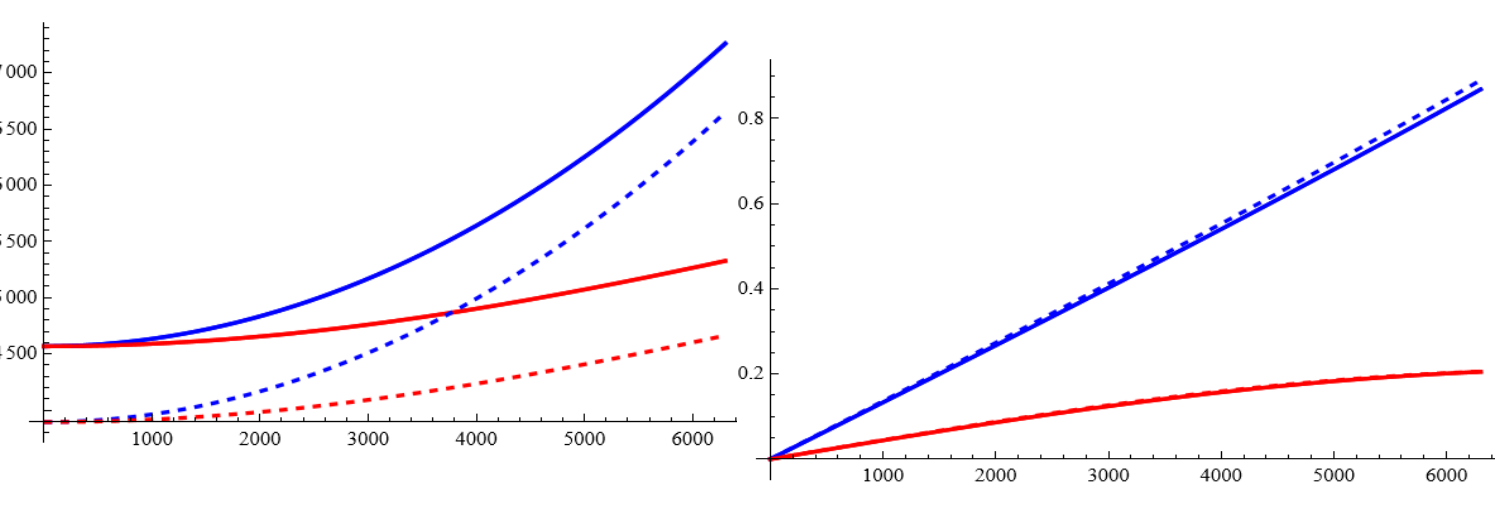

Figure 23: (a) Sagittal (blue) and tangential (red) radii of curvature of M1 vs radius; (b) derivative of the RoC with radius. In both cases, the solid curves are for A27, the dashed lines are for B25. The A27 curves are drawn both with and without the aconic term; the maximum difference amounts to 0.04%, much too small to see in these plots.

**A27 vs B25 survey speeds**

We are now in a position to compare the overall survey speeds for the two designs. As argued in Section 4, it is the performance at a field radius 0.866° that determines the overall survey speed. Based on the difference in vignetting and *ZIQ* (0.214" vs 0.321") it is then straightforward to calculate the relative survey speeds, as shown in Figure 24. The B25 design gives an optimal survey speed 9% slower, with an optimal fiber size 9% larger, than A27. The combined penalty on the speed per unit spectrograph cost is ~35%. These numbers do not alter much if average *ZIQ* over positioners is used instead of the *ZIQ* at 0.866°.

The B25 design has optimal fiber size ~ 1.0", which is also the baseline fiber size. That is, the nominal *FIQ* 'allowance' is used up by the *ZIQ*. But all the other *FIQ* contributions are not yet included, so it is obvious that the fiber size will be undersized for the B25 design. The A27 design has optimal fiber size 0.92", retaining some modest headroom (~0.24") for non-*ZIQ* factors.

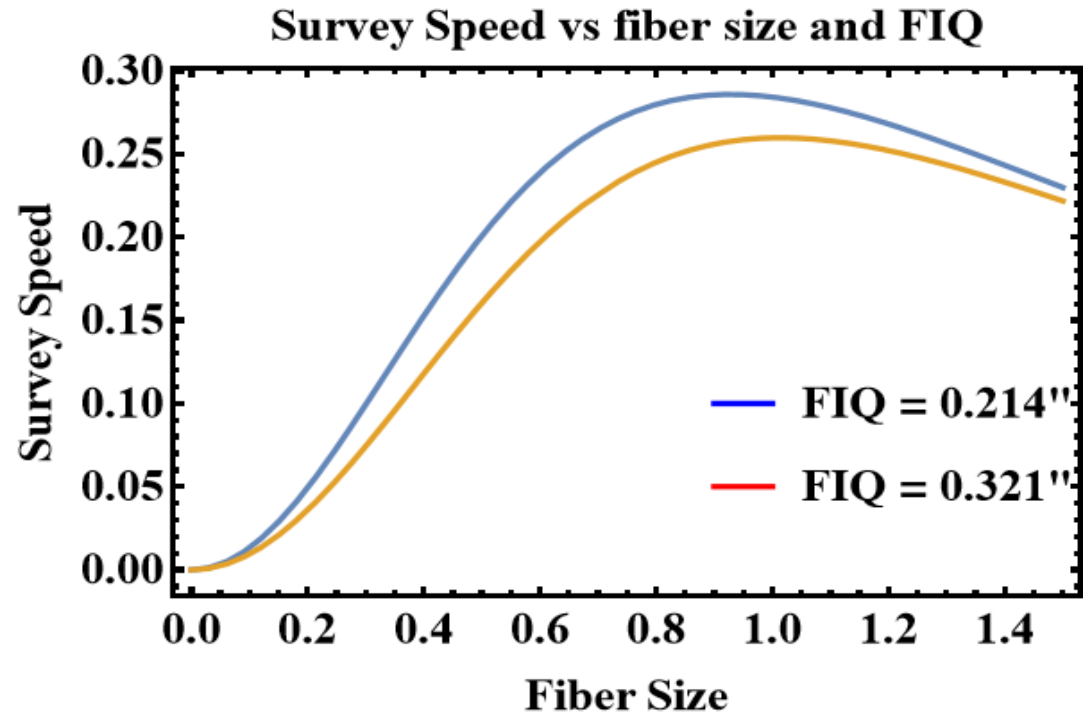

Figure 24. Average survey speed vs fiber size, for face-on $R_e$=0.1" galaxies, for B25 vs A27 designs, based on their vignetting and *IQ* at 0.866°. The differences at the maxima are a decrease in survey speed of 9%, and an increase in the fiber diameter of 9% (1.0" vs 0.92").

## [7] IFS EVOLUTION

The IFS train as presented at SPIE 2024 [4] is shown in Figure 4. It is convoluted, and also suffers from 20% IFS vignetting due to the M3/M4 pickoff.

Multiple other variants have been investigated subsequently (Figure 25). A significant advance was to move the sub-field selection to M3 ( despite some doubts over doing this at mirror almost at focus). This means that M4 needs only be sized for the IFS sub-field (initially 5' radius), and could hence (just) be accommodated within the shadow of the top-end in the return beam (Figure 25a). This reduces the mirror count significantly. Two options were initially explored, with the beam either passing directly downward through the hole in M1, or (Figure 25a) passing via Nasmyth focus. Both options have issues: the former option involves removing at least 2 and more likely 3 M1 segments, with consequent light-loss; also the beam (which is invariant with elevation) passes at the side of the tower structure supporting the WFC (which does move), so the tower is poorly supported laterally, raising issues for earthquake resistance. Also, the required meter-class collimator and fold mirrors (M5, M6) interfere with the phasing of the inner M1 segments. Finally, the area around the focal surface is extremely congested already, and there was deemed to not be enough space for the beam, given all the required cable wraps etc. Given that the fibers are now proposed to exit the telescope directly via a suspended helical cable, this issue might merit revisiting.

The latter option (Figure 25a) requires a very slow (~F/50 ) beam to reach the desired IFS focus, meaning large K-mirrors, and it is desired that these be rotatable rather frequently (every few minutes), so there were concerns about movement and settling times. The GLAO mirror position and AOI is also not optimal.

Subsequently, the IFS FoV was increased to 6', to increase NGS sky coverage. This reintroduced vignetting due to M4. Hence the adopted solution (Figure 25b) has an expanding beam coming from M5, eliminating the vignetting and allowing a much faster final beam speed (F/28.5), at the cost of additional large optics at Nasmyth and somewhat increased conflict with MOS through M3 height and guide camera clearance.

Note that all solutions require at least two transmissive windows (which can be powered): one to seal the positioner volume against humidity and dust, and one to thermally seal the telescope from the IFS spectrograph room, since the spectrographs will operate at fixed temperature year-round.

The current baseline design (25b) is discussed in more detail in D26 [2].

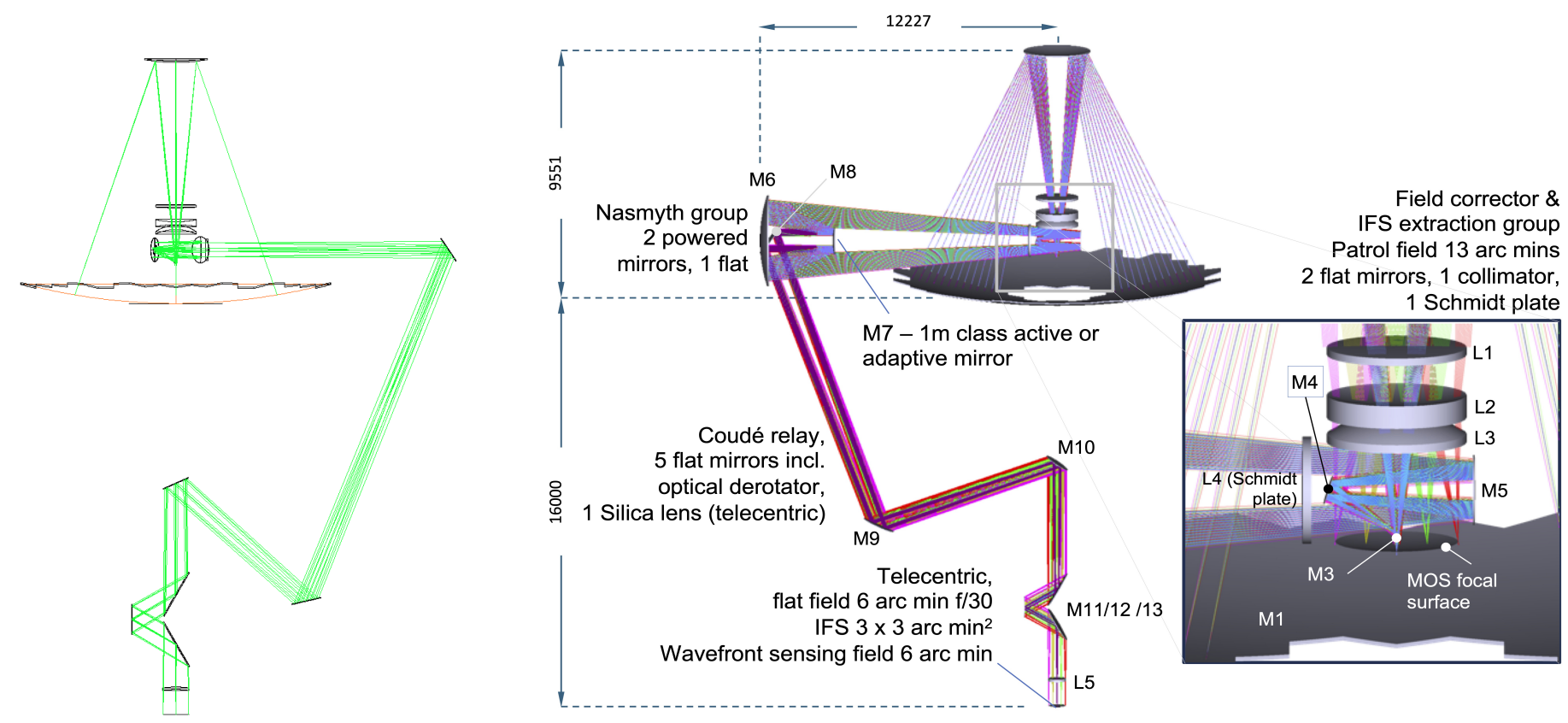

Figure 25(a):. Initial IFS layout with sub-field pickoff at M3, and very slow (F/50) focus ; (b) the current baseline design [2]. The additional optics at Nasmyth allow a deformable mirror with excellent size, AOI and conjugation, and giving a more tractable (F28.5) final focus.

## [8] CONCLUSIONS

Viable WST designs have been developed, despite the challenges presented by the enormous MOS etendue and the requirement for simultaneous MOS and IFS use. The solution adopted as the MOS baseline design has the lowest risk, but does not have the best as-designed optical performance. We have argued that optical performance is very important for a massively MOS telescope, dramatically impacting both survey speed and spectrograph costs. The obstacles to adopting a design with better optical performance are technical: it must be demonstrated that
(a) A larger M2 can be stiff enough to allow the fast-steering required to correct for wind-shake, or alternatively be multiply supported (eg by pneumatic or hybrid actuators);
(b) A very slightly aconic M1 can be manufactured and the segments controlled in shape and precision to the required level;
(c) M1 can be adequately phased without full visibility of the entire hexagonal shape of every segment - either because they have adequate intrinsic stiffness, or by physically trimming the unwanted corners. Smaller segment size would also help the tension between phasing and IQ requirements, by making both outer and inner edges of M1 rounder.

## ACKNOWLEDGEMENTS

Bernard Delabre first showed how IFS and MOS could be combined in a single telescope. Paulo Spano alerted us to the 'effectivly-pupil-centric' nature of the PFS design. The loss-less ADC design would never have progressed as it has without the input of Peter Gillingham, whose creativity, talent, and humour are all greatly missed.

## APPENDIX: ATMOSPHERIC REFRACTION FOR WIDE-FIELD TELESCOPES IN ZEMAX

Bizarrely, Zemax does not deal correctly with atmospheric refraction for wide-field telescopes, because the same refraction is applied at all field positions. A simple physical model of the Earth's atmosphere has been implemented for all the designs presented here, which seems to do an adequate job (for natural seeing telescopes) of reproducing tabulated atmospheric refraction values. The model is that the atmosphere consists of a shell of air of uniform density equal to that at ground level, and of a thickness equal to its scale height. A sketch is shown in Figure 26. The tangent plane at the point of entry of the rays into the atmosphere is tilted from horizontal by an amount $\Delta Z = \arcsin[\sin(ZD)\, R/(R+H)]$ where $R$ is the earth's radius (6377km for Paranal at 24°N) and $H$ is the scale height (assumed to be 8km). Further refinements (oblate ellipse for the atmosphere, scale height or humidity varying with altitude) are possible but do not seem necessary. It would be attractive to simply include the Earth and atmosphere directly as optical elements in Zemax as was initially tried, but this leads (unsurprisingly!) to ray-trace errors.

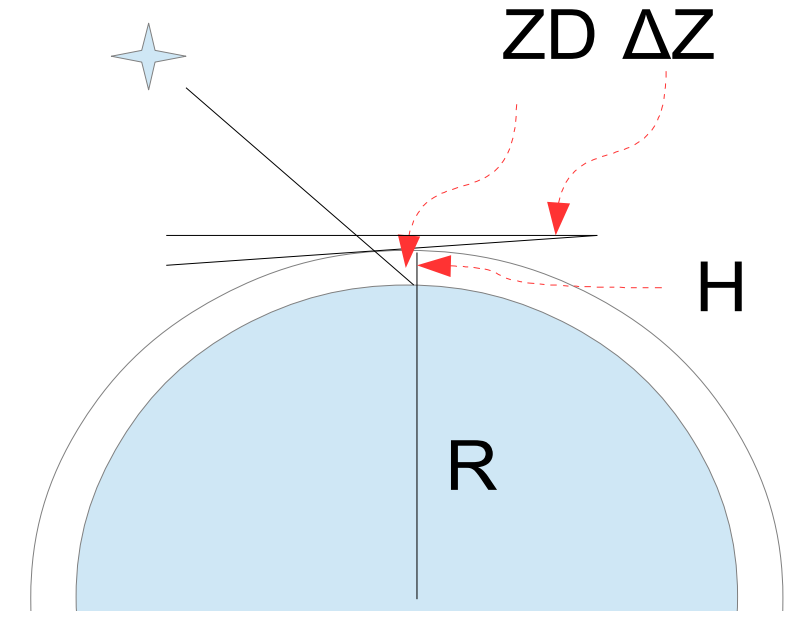

Figure 26: Sketch of the model for atmospheric refraction. The Earth's radius is R, the atmosphere is assumed to have constant density as at ground level, with a thickness equla to the scale height H. The Zenith Distance is Z, the tangent plane at ray entry into the atmosphere is tilted wrt horizontal by ΔZ, reducing the apparent Zenith Distance.